# POTENTIAL FOR LASER-INDUCED MICROBUNCHING STUDIES WITH THE 3-MHZ-RATE ELECTRON BEAMS AT ASTA*

A.H. Lumpkin[#] and J. Ruan, Fermi National Accelerator Laboratory, Batavia, IL 60510 USA
J.M. Byrd and R.B. Wilcox, Lawrence Berkeley National Laboratory, Berkeley, CA 94720 USA

*Abstract*

Investigations of the laser-induced microbunching as it is related to time-sliced electron-beam diagnostics and high-gain-harmonic generation (HGHG) free-electron lasers using bright electron beams are proposed for the ASTA facility. Initial tests at 40-50 MeV with an amplified 800-nm seed laser beam co-propagating with the electron beam through a short undulator (or modulator) tuned for the resonance condition followed by transport through a subsequent chicane will result in energy modulation and z-density modulation (microbunching), respectively. The latter microbunching will result in generation of coherent optical or UV transition radiation (COTR, CUVTR) at a metal converter screen which can reveal slice beam size, centroid, and energy spread. Additionally, direct assessment of the microbunching factors related to HGHG by measurement of the COTR intensity and harmonic content after the chicane as a function of seed laser power and beam parameters will be done. These experiments will be performed using the ASTA 3-MHz-rate micropulse train for up to 1ms which is unique to test facilities in the USA.

## INTRODUCTION

We have identified critical aspects of laser-induced microbunching (LIM) to be explored at the Advanced Superconducting Test Accelerator (ASTA) which relate directly to time-sliced electron-beam diagnostics and seeded free-electron laser (FEL) issues. In the first category the capability of evaluating the electron beam parameters such as beam size, centroid, and energy spread as well as the microbunching factors in sub-ps time slices should be possible by imaging the LIM coherent optical and UV transition radiation (COTR, CUVTR). In the second category, enhanced performance in gain length, spectral bandwidth, central wavelength stability, etc. of free-electron lasers (FELs) can be obtained by seeding the FEL either with electron beam microbunching at the resonant wavelength as in the case of high gain harmonic generation (HGHG), cascaded HGHG, and echo-enhanced harmonic generation (EEHG) [1-3] or by generating a photon beam at short wavelengths such as from a plasma as in high harmonic generation (HHG) [4]. These seeding processes are initiated by co-propagating a laser beam with the electron beam through a short undulator (modulator) tuned to the seed wavelength to generate an energy modulation which is then converted to a z density modulation (microbunching) in a dispersive section such as a chicane. It has been found that the electron beam is also microbunched at the harmonics of the laser fundamental after the chicane, and the radiator undulator is tuned to one of them for the FEL [1]. Yu et al. have reported lasing on the 3[rd] harmonic of the laser fundamental of 800 nm, and FERMI@Elettra staff have reported recent HGHG results out to the 13[th] harmonic of 266 nm [5]. The direct measurement of the microbunching by looking at the coherent optical and UV transition radiation (COTR) was first done in a SASE FEL [6], but microbunching generated by LIM has rarely been measured [7]. Moreover, high harmonic content has not been directly observed, but it has been deduced from the HGHG radiator results [8]. Direct microbunching measurements have been proposed on the SDUV FEL in Shanghai [9], and discussions are underway with FERMI@elettra staff for VUV tests on their HGHG FEL. Elucidating the harmonic content, optimizing it, and benchmarking codes would be critical to present and future short wavelength (VUV-soft x-ray) FEL projects based on HGHG or EEHG. Only ASTA has the high-micropulse-repetition rate (3 MHz for 1 ms) beam such as proposed for the next generation of FELs [10], albeit with higher duty factor for the latter.

Our emphasis in this paper is on the use of COTR (ultimately CVUVTR) as a direct microbunching diagnostic with the potential for time–resolved electron beam diagnostics and for benchmarking relevant FEL codes. As context, we point out that there are at least three mechanisms for generating optical-regime microbunching (which can be extended to the VUV) in an electron beam:

1. longitudinal-space-charge-induced microbunching (LSCIM): This mechanism has been suggested by Saldin et al. [11] to contribute more in the visible wavelength regime than coherent synchrotron radiation (CSR) or wakefields. This effect can be considered as starting from shot noise in the charge distribution that couples through the longitudinal impedance of the transport line and linac into an energy modulation. This energy modulation will become a z-density modulation, or microbunching, after bunch compression in a chicane (or other $R_{56}$ lattice point) [12]. It is generally a broadband effect in wavelength, and most of the gain is in the FIR (>10-µm regime). However, we have the LCLS/SLAC [13], APS/ANL [14], FLASH/DESY [15], and FERMI@Elettra [16] results on COTR in the OTR images and even scintillator screens. Spatially localized enhancements of 10-10,000 at visible wavelengths have been reported which prevent using the standard beam profiling techniques with optical beam images.

___________________
[#] lumpkin@fnal.gov
*Work supported under Contract No. DE-AC02-07CH11359 with the United States Department of Energy.

2. laser-induced microbunching (LIM): a) where an external laser beam is injected into the beamline so it copropagates with the electron beam through a short undulator (the modulator) which interaction modulates the beam energy that then becomes a z-density modulation after a dispersive element such as in a chicane and b) where the seeding of an FEL with an external laser or harmonic at the long radiator undulator itself. The former results in narrowband microbunching and is used to prebunch or seed the beam for the FEL, and the latter is used as input to an FEL amplifier. The main objective of these proposed studies is measuring LIM.

3. SASE-induced microbunching (SIM) which is the fundamental mechanism of the self-amplified spontaneous emission (SASE) FEL process starts from noise when the SASE photon fields acting with the undulator fields on the electrons result in the growth of electron beam microbunching and concomitant exponential growth of SASE light at the resonant wavelength and harmonics. This is a narrowband effect. We note evidence for the LSCIM from the linac providing prebunched beam on experiments in the APS visible SASE FEL [17]. This demonstration actually links to the HGHG process via the principle of a prebunched beam's enabling FEL startup, as opposed to startup from noise as in a SASE FEL. Historically, LSCIM [18] and SIM COTR [19] were modeled after the first experiments, but in the LIM case we have more modeling in place before extensive experiments have been performed. We also expect to take advantage of techniques developed for the earlier modes.

## CONCEPTUAL ASPECTS of LIM

### Laser-Induced Microbunching

The initial step in HGHG is co-propagation of a seed laser with the electron beam through a short undulator or modulator tuned to resonance as indicated schematically in Fig. 1. The laser pulse length can be used to modulate a time slice of the transverse distribution or it could be lengthened to provide modulation over the whole e-beam pulse length. At 40 MeV it is impractical to satisfy the resonance condition for an 800-nm wavelength on the fundamental of an undulator based on permanent magnets. For a planar undulator, the FEL process is governed by the resonance condition:

$$\lambda = \lambda_u (1 + K^2/2)/2n\gamma^2, \qquad \text{Eq. 1}$$

where $\lambda$ is the FEL wavelength, $\lambda_u$ is the undulator period, $K$ is the undulator field strength parameter, n is the harmonic number, and $\gamma$ is the relativistic Lorentz factor. It has been calculated that there is reasonable coupling strength at the third harmonic of the planar undulators as shown in Fig. 2 from reference [20]. The coupling coefficients $JJ_n$ for n=1,3,5 versus $K$ are reasonable, although with a rapid decrease in their values below $K=1.5$ for n=3,5 for the 780-nm case. We suggest with this concept we could modulate the energy at 800 nm in an undulator with period 2.18 cm using the third harmonic coupling of ~0.25 and then generate harmonics of the laser wavelength in the microbunching in the dispersive section. Initially, we would use visible-UV optics to detect n=2,3,4 at 400, 266, and 200 nm, respectively, in the Phase 1a experiments. Subsequently we would look for the critical higher harmonics in the VUV with appropriate diagnostics in Phase 1b as will be discussed in a later section. (Due to available 1054-nm lasers now at FNAL and LBNL, there may be some practical advantages for shifting to this wavelength.

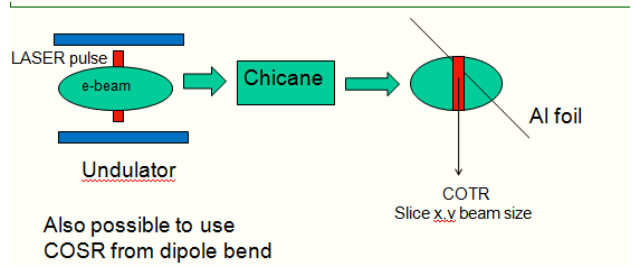

Figure 1: Schematic of laser-induced energy modulation followed by the dispersion in the chicane to produce a microbunched beam that can generate COTR or COSR in a slice. In principle the entire e-beam profile could be microbunched as well for HGHG.

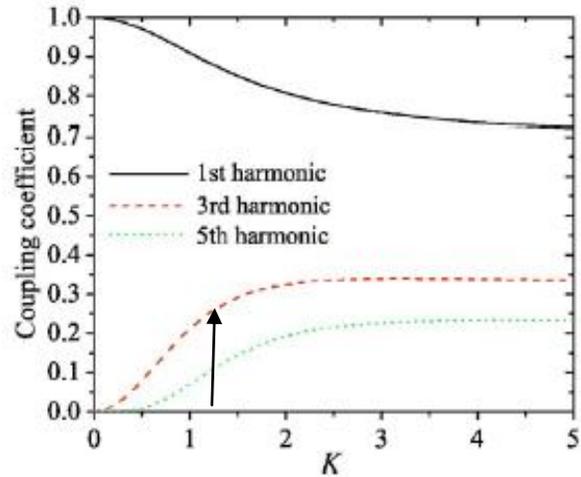

Figure 2: Coupling coefficients for odd harmonics for a planar wiggler as a function of K. The $\gamma$=90 case with K=1.20 is indicated by the arrow. [20]

However, the present S20 PC streak camera tube is insensitive at 1054 nm, it is difficult to chirp the IR beam to generate few-ps pulses to cover the electron beam longitudinally, and the harmonic wavelengths will be 20% longer than the reference case. Further evaluation of this aspect is warranted. The narrowband nature of LIM may be best illustrated by showing the results of the narrowband microbunching in the SIM experiments at

APS/ANL in 2002. Figure 3 shows a composite of the spectra obtained for a visible FEL operating near 535 nm [6]. The COTR spectra on the left are narrowband, but not quite as narrow as the SASE also observed after undulator 5 of nine undulators. The bunching fraction at saturation was calculated at about 20% of the e-beam, and enhancements of $10^4$ for COTR and $10^6$ for SASE were obtained. As we understand it bunching fractions of 5-10 % could be obtained in HGHG and EEHG configurations, even for harmonics of n=10 or 50, respectively. The potential for measuring the resultant enhanced transition radiation and elucidating LIM is high. We note that the bunching fractions would also apply to enhancing coherent optical synchrotron radiation (COSR), coherent undulator radiation (CUR), and coherent optical diffraction radiation (CODR) which mechanisms provide the potential for nonintercepting beam diagnostics for the high power beam mode of ASTA.

installed in the form of three cryomodules with eight 9-cell cavities with average gradient of 31 MV/m after the chicane. The phase of the CC2 section can be adjusted to energy chirp the beam entering the chicane to vary bunch-length compression. Maximizing the FIR coherent transition radiation (CTR) in a detector after the chicane will be used as the signature of generating the shortest bunch lengths. Micropulse charges of 20 to 3200 pC will be used typically. The nominal pulse format for high power ILC-like beam is 3.2 nC per micropulse at 3 MHz for 1 ms. The macropulse repetition rate will be 5 Hz. For the 1-ms period the pulse train micropulse spacing can be at a higher rate than a proposal of 100 kHz in any one of ten FEL beamlines. This aspect is unique for test facilities in the USA and highly relevant to the next generation of FELs..

We would start Phase 1 at 40-50 MeV with an 800-nm seed and a modulator with about 1.8 to 2.2 cm period, K~1.2, and with length about 1 m. The experiment location is shown in Fig. 4. We would need to use resonance at the 3rd harmonic of the planar undulator as explained earlier and shown in Table 1. This is a key, and confirmation of this approach and modeling of the expected microbunching is recommended. We would need an amplifier for the 100 kHz 800-nm laser beam from the ASTA Ti:Sa laser or doubled- frequency Erbium Fiber laser to provide about 200-500 μJ per pulse, although we could start with fewer pulses in the pulse train initially. An optical parametric amplifier (OPA) could do this in principle. A laser concept developed for the electro-optical sampling (EOS) experiments envisioned for ASTA is shown in Fig. 5. The pump laser is an existing Nd:YLF operating at 1 MHz. One might obtain higher energy pulses at the lower repetition frequency of 100 kHz and for longer pulses. The pulse train also might be shortened for initial tests. The modulator should fit after the dogleg entering the low-energy test beamline at ASTA with diagnostics at both ends to align the laser and electron beam transversely. The FNAL Visible-UV streak camera would be used for timing the laser with the UR or OTR within

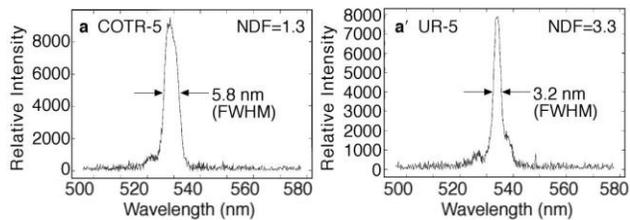

Figure 3: Example COTR (L) and SASE (R) spectra from the visible FEL experiments at ANL sampled after undulator 5 with SASE saturation occurring at about undulator 5. The SASE is 100 times brighter than COTR [6].

## PROPOSED STUDIES

The ASTA linac with photocathode (PC) rf gun, two booster L-band SCRF accelerators (CC1 and CC2), and beamline is schematically shown in Fig. 4. The L-band accelerating sections will provide 40- to 50-MeV beams before the chicane, and an additional acceleration capability up to a total of 800 MeV will eventually be the modulator to 1 ps or better. We have successfully done this in recent EOS tests at A0PI with the 800 nm Ti:Sa and incoherent OTR signals [21]. It also might be used to look directly at the LIM COTR within the micropulse as generated at the test station. (The system temporal resolution will be improved with narrow bandwidth COTR and also might be extended with a deflecting mode cavity [7].) The chicane configuration that is planned for an emittance exchange (EEX) could presumably be used with its tunable $R_{56}$ [22], but this needs to be checked. Visible-UV diagnostics would be needed after the Chicane to measure the harmonics n=2,3,4 of 800 nm at 400, 266, and 200 nm, respectively. In Phase 1b we would use VUV diagnostics since codes generally indicate harmonics of n~10 should be generated.

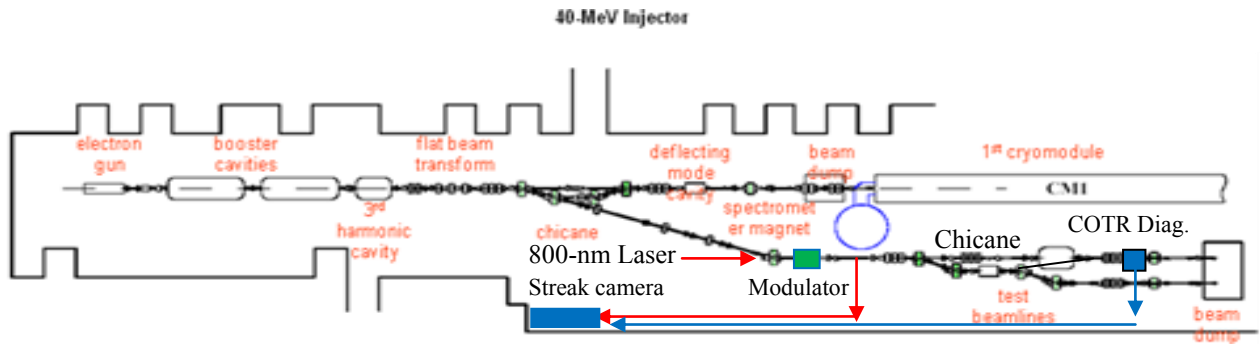

Figure 4: Schematic of the injector for the ASTA facility showing PC rf gun, booster accelerators, and beamlines. The first LIM tests are proposed in the test beamline as schematically indicated by the 800-nm laser beam injection into the beamline, modulator, chicane, and COTR diagnostics station locations. The Visible-UV streak camera could be used for both laser synchronization and COTR measurements.

Table 1: Summary of possible microbunching experiments at ASTA at 40-45 MeV and higher energies following the cryomodules. The potential observation wavelengths of several of the harmonics are indicated.

| Phase # | Energy (MeV) | Laser Fund. (nm) | Und. Period (cm),K,n | Harm. (nm) |
|---|---|---|---|---|
| 1a | 44.5 | 800 | 2.18,1.2,3 | 400,266,200 |
| 1b | 44.5 | 800 | 2.18,1.2,3 | 100,90,80 |
| 2 | 250 | 800 | 20.0,1.36,1 | 400,266,200 |
| 2 | 200 | 266 | 5,1.2,1 | 48,29,26 |

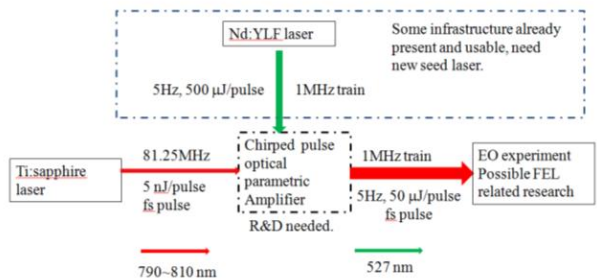

Figure 5: Schematic of the proposed ASTA diagnostics laser with OPA indicated. The pulse length and wavelengths need to be evaluated for microbunching tests. This laser system is being considered for the downstream diagnostics lab to support EOS tests.

## SUMMARY

In summary, we have described how a series of direct LIM experiments with diagnostics in the optical, VUV, and soft x-ray regime could be implemented at the ASTA facility. This is the only facility in the USA that has the pulse train of at least 100 kHz repetition rate to simulate proposed FEL configurations. These experiments should extend understanding of the critical phenomena of microbunching harmonic generation and preservation and allow benchmarking of codes. In addition, the as-described sub-micropulse electron-beam diagnostics based on COTR could be developed into nonintercepting electron beam diagnostics if used with the COSR, CUR, or CODR mechanisms.

## ACKNOWLEDGMENTS

The FNAL authors acknowledge discussions with M. Wendt, M. Church, V. Shiltsev, S. Nagaitsev, and S. Henderson of FNAL on the ASTA facility and AARD program.